\pdfoutput=1
\documentclass[aps,prc,reprint,groupedaddress,nofootinbib]{revtex4-1}
\usepackage[utf8x]{inputenc}

\usepackage{amssymb}
\usepackage{amsmath}
\usepackage[colorlinks=true,linkcolor=blue,citecolor=blue, urlcolor=blue]{hyperref}   
\usepackage{tikz}
\usepackage{graphicx,mathrsfs}

\def\Eq#1{Eq.~(\ref{#1})}
\def\Eqs#1{Eqs.~(\ref{#1})}
\def\eq#1{(\ref{#1})}
\def\app#1{Appendix~\ref{#1}}
\def\Fig#1{Fig.~\ref{#1}}

\def\Sec#1{Sec.~\ref{#1}}

\def\p{{\bf p}}
\def\q{{\bf q}}

\def\k{{\bf k}}

\begin{document}

\title{Fast resonance decays in nuclear collisions}
\author{Aleksas Mazeliauskas}
\email[]{a.mazeliauskas@thphys.uni-heidelberg.de}
\affiliation{Institut f\"{u}r Theoretische Physik, Universit\"{a}t Heidelberg, 
69120 Heidelberg, Germany}

\author{Stefan Floerchinger}
\email[]{stefan.floerchinger@thphys.uni-heidelberg.de}
\affiliation{Institut f\"{u}r Theoretische Physik, Universit\"{a}t Heidelberg, 69120 Heidelberg, Germany}

\author{Eduardo Grossi}
\email[]{e.grossi@thphys.uni-heidelberg.de}
\affiliation{Institut f\"{u}r Theoretische Physik, Universit\"{a}t Heidelberg, 69120 Heidelberg, Germany}

\author{Derek Teaney}
\email[]{derek.teaney@stonybrook.edu}
\affiliation{Department of Physics and Astronomy, Stony Brook University, Stony 
	Brook, NY 11794, USA}

\date{\today}

%
\begin{abstract}
 In the context of ultra-relativistic nuclear collisions, 
 we present a fast method 
 for calculating the final particle spectra 
 after the direct decay of resonances from a Cooper-Frye integral over the  freeze-out surface.
 The method is based on identifying 
 components of the final particle spectrum that transform in an irreducible way under rotations in the fluid-restframe. Corresponding distribution functions can be pre-computed including all resonance decays. 
 Just a few of easily tabulated scalar functions then
 determine the Lorentz invariant decay spectrum from each space-time point, and
 simple integrals of these scalar functions over the freeze-out surface 
 determine the final decay products.  
 This by-passes numerically costly event-by-event calculations of the intermediate resonances.
 The method is of considerable practical use for making realistic
  data to model comparisons of the identified particle yields and flow harmonics, and for studying the viscous corrections to the freeze-out distribution function. 
%

\end{abstract} 
\maketitle
\section{Introduction\label{sec:introduction}}

Ultra-relativistic heavy ion collisions create the deconfined state of matter called the Quark-Gluon Plasma (QGP), which has been under intensive experimental and theoretical research in the last two decades~\cite{Busza:2018rrf,Heinz:2000ba}. Remarkably, the expanding QGP  has been very successfully described as a relativistic fluid, where the system dynamics is completely determined by a few macroscopic fields like  fluid velocity $u^\mu(x)$ or temperature $T(x)$~\cite{Heinz:2013th,Teaney:2009qa,Luzum:2013yya,Gale:2013da,deSouza:2015ena}. As the fluid expands and cools down below the cross-over temperature $T_c\approx 155\,\text{MeV}$, quarks and gluons are re-confined in hadronic degrees of freedom.  Therefore a systematic comparison between the hydrodynamic models of the QGP and experimental data necessitates the conversion of hydrodynamic fields into hadronic degrees of freedom.

Various techniques of treating the hadronic phase have been developed over the years. Resonances are sampled at the freeze-out surface using the Cooper-Frye formula~\cite{Cooper:1974mv} and then passed to hadronic transport models, which describe both the decays and possible rescatterings of resonances~\cite{Bass:1998ca,Petersen:2018jag}. 
However direct resonance decays (without rescatterings) are often used in phenomenological studies~\cite{Alba:2017hhe,Eskola:2017bup,Niemi:2015qia,Bozek:2013ska}. The decay processes of resonances are simulated by Monte-Carlo generators~\cite{Torrieri:2004zz,Amelin:2006qe,Chojnacki:2011hb}, or by semi-analytic treatments of decay integrals~\cite{Sollfrank:1991xm,Sollfrank:1990qz}.
From  $\sim 300$ species of hadronic  resonances produced in high energy nuclear collisions, only a handful long-lived hadrons (e.g. pion, kaons and protons) reach the  particle detectors and are directly observed~\cite{Abelev:2012wca,Tanabashi:2018oca}.
In this paper we show how to by-pass the numerically costly procedure of calculating the intermediate resonance decays and to relate directly the hydrodynamic fields on the freeze-out surface to the final decay particle spectra.

Let us remark here that semi-analytic description of resonance decays was studied previously~\cite{Sollfrank:1991xm,Sollfrank:1990qz,Becattini:2001fg,Broniowski:2001uk}.
While these works constitute the basis of our approach, our framework is applicable to arbitrary freeze-out surfaces and more general particle distribution functions.

In \Sec{sec:map} we describe a particle  decay process as a linear Lorentz invariant map, which transforms the  spectrum of initial (or primary) particles to the spectrum of final decay products. In \Sec{sec:cooperfry} we argue that the decay map can be applied directly to the particle distribution function \emph{before} performing the Cooper-Frye integral and we define a distribution function for the decay products. Using group theoretical arguments we find the Lorentz invariant decomposition of the decay particle distribution functions as a sum of frame-independent \emph{weight functions}, which we calculate.
 In \Sec{sec:perturbations} we show that the same procedure also applies to viscous and linear perturbations of  particle distribution function on the freeze-out hypersurface. Then in \Sec{sec:modebymode} we discuss the implementation of fast resonance decay procedure for general freeze-out surfaces and phenomenologically convenient setups of blast-wave freeze-out and  mode-by-mode hydrodynamics.
We end with discussion of further extensions and applications in \Sec{sec:conclusions}. Finally \app{sec:decomp} gives the derivation  of the irreducible decay spectrum components.

\section{Lorentz invariant decay map\label{sec:map}}

Relativistic particle decays is a well established subject~\cite{Byckling:1971vca,DeGroot:1980dk,Landau:1982dva,Tanabashi:2018oca,Sollfrank:1991xm,Sollfrank:1990qz} and here we briefly discuss some of the key formulas.
In kinetic theory, decays can be treated as $1\leftrightarrow n$ particle scatterings. The probability for such an event is given by a Lorentz invariant integral over the scattering matrix squared $|\mathcal M|^2$, the available momentum phase space (constrained by 4-momentum conservation) and the phase space densities of initial and final particles, i.e.\ the gain and loss terms.
 In chemical and thermal equilibrium both the decay and the reverse process are equally likely, however, if the system becomes dilute and falls out of the detailed balance,  multi-particle scatterings become very improbable and the $1\rightarrow n$ decays, which are \emph{linear} in the initial particle densities, dominates the subsequent phase space evolution of particles. This  is exactly what happens for hadron resonances in heavy ion collisions below the freeze-out temperature. At sufficiently late times, all allowed decays will have taken place and the Lorentz invariant  spectrum of the final particle species $b$ will be proportional to the primary populations of resonance species $a$, which decay (directly or through intermediate resonances) to particles of type $b$\footnote{This applies to all strong decays of hadron resonances, which have typical lifetimes $\tau \sim 10^{-23}\,\text{s}$ and decay before reaching the detector. Some weak decays of strange particles take place within the detector and can be reconstructed experimentally~\cite{Aamodt:2011zj,Kalweit:2011if}.}.

A particle decay is intrinsically a probabilistic process, and the resultant particle spectrum from a decay cascade will fluctuate event by event, but for  very large number of initial resonances (or an average over many decay cascades), we can write the $1$-body particle spectra of final particles as a
 Lorentz invariant integral over the primary resonances\footnote{We also note here that the two-particle distribution $E_\p E_\q dN_{bc}/(d^3\p d^3\q)$ inherits correlations from a decay $a\rightarrow b+c$ as a consequence of energy and momentum conservation. However, in the present paper we only concentrate on one-particle distributions.}
\begin{equation}
E_\p \frac{d N_b}{d^3 \p} =  \int\! \frac{d^3 \q}{(2\pi)^3 2 E_\q}  D^a_b(\p,\q) \; E_\q \frac{d N_a}{d^3 \q}.\label{eq:map}
\end{equation}
The linear decay map $D^a_b(\p,\q)$ simply gives the Lorentz invariant probability of particle $a$ with momentum $\q$ to decay to a particle $b$ with corresponding momentum $\p$.
  Summing over all species of primary resonances $a$ then gives the \emph{total decayed particle spectrum} of particle species $b$. 
 We note in passing that the decay map $D_b^a(\p,\q)$ fulfils certain sum rules as a consequence of conservation laws for energy, momentum, net baryon number, electric charge, etc. 

In general the linear map $D^a_b(\p,\q)$ is a composition of phase space integrals, 4-momentum conservation and decay matrix elements for each successive decay in the decay cascade~\cite{Sollfrank:1990qz,doi:10.1142/5427}. Most of the listed decays or hadron resonances in the Particle Data Group book~\cite{Tanabashi:2018oca} are 2-body and 3-body decays, which in heavy ion  simulations are customary approximated as isotropic decays with a branching ratio $B$.
For the simple case of \emph{isotropic} two-body decay $a\rightarrow b+c$ the  phase space integral of the decay partner $c$ can be done analytically and the map $D^a_{b|c}$ is reduced  to a Lorentz invariant delta function of the  product of initial and final particle momenta  $p^\mu q_\mu$~\cite{Sollfrank:1990qz,Byckling:1971vca,Tanabashi:2018oca} 
\begin{align}
D^a_{b|c}(p^\mu q_\mu)
&=B\frac{4\pi^2 m_a}{p^a_{b|c}} \delta( q^\mu p_\mu+ m_a E^a_{b|c}),\label{eq:map2body}
\end{align}
where $B$ is the branching ratio for this process. In the rest-frame of particle $a$, \Eq{eq:map2body} is  simply a uniform probability distribution   on a sphere with radius $|\p|=p^{a}_{b|c}$ fixed by  energy conservation
\begin{equation}
p^a_{b|c} \equiv
\frac{1}{2m_a} \sqrt{((m_a+m_b)^2-m_c^2)((m_a-m_b)^2-m_c^2})\label{eq:pstar},
\end{equation}
and  we also use $E^a_{b|c}\equiv\sqrt{m_b^2+(p^a_{b|c})^2}$. 

Isotropic three body decays $a\rightarrow b+c+d$  have larger phase-space and 4-momentum conservation is not enough to fix the particles' momenta even in the rest-frame of the primary resonance $a$. However treating the two partner particles $c$ and $d$ as a fictitious particle $\tilde c$ with an effective mass $m_{\tilde c}^2=-(p_c+p_d)^2$, the three body decay map $D_{b|cd}^a$ can be written as an integral of the  2-body decay map for the allowed values of $m_{\tilde c}$~\cite{Sollfrank:1990qz,Byckling:1971vca,Tanabashi:2018oca}
\begin{equation}
D^a_{b|cd}(p^\mu q_\mu)= \frac{\int_{m_c+m_d}^{m_a-m_b}\!\!dm_{\tilde c}\,\, p^{a}_{b|\tilde c} p^{\tilde c}_{c| d}  D^a_{b|\tilde c}(p^\mu q_\mu)}{\int_{m_c+m_d}^{m_a-m_b}\!\!dm_{\tilde c}\,\, p^{a}_{b|\tilde c} p^{\tilde c}_{c|d}}\label{eq:3body}.
\end{equation}
where $p^{a}_{b|\tilde c}$ is the momentum of $b$ in the rest-frame of $a$ and $p^{\tilde c}_{c|d}$ is the momentum of particle $c$ or $d$ in their common rest-frame~\cite{Tanabashi:2018oca}.

Composing 2-body and 3-body decay maps, \Eq{eq:map2body} and \Eq{eq:3body}, according to the chain rule
\begin{align}
&D^a_{b}(p^\mu q_\mu)=\int \frac{d^3\k}{(2\pi)^32 E_\k}D^e_{b}(p^\mu k_\mu)D^a_{e}(k^\nu q_\nu)\label{eq:chain}
\end{align}
a map for an arbitrary decay cascade can be constructed.
Importantly, this $a\rightarrow b+X$ decay map will be itself only a function of the invariant product $p^\mu q_\mu$ of initial and final particle momenta.
 Such a map can be calculated iteratively by applying the 2-body decay map \Eq{eq:map2body} for which the chain rule \Eq{eq:chain} can be reduced to just a single dimensional integral over dimensionless variable $w$\footnote{In the rest-frame of the particle $b$ the variable $w$ has the physical interpretation as the fraction of  particle's $a$ momentum in the direction of the fluid velocity. The limit of the massless final state particle $m_b=0$ can be treated by a simple change of variables $u=(1-w)\frac{m_e^2}{m_b^2}$.}
\begin{align}
D^a_{b}(p^\mu q_\mu)=
\label{eq:mapmap}
B \frac{1}{2}& \frac{ m_e^2}{m_b^2}\nonumber
\int_{-1}^1\! dw\,
 D^a_e\Big(q^\mu p_\mu \frac{m_e E^{e}_{b|c}}{m_b^2}\\
&+ w\sqrt{(q^\mu p_\mu)^2-m_a^2 m_b^2} \frac{m_e p^e_{b|c}}{m_b^2} \Big)
\end{align}
The extension to three body decays follows immediately by the application of \Eq{eq:3body}.

The decay chain map $D^a_{b}(q^\mu p_\mu)$ is independent of initial particle spectrum and  only depends on particle properties and branching ratios listed in the particle data book~\cite{Tanabashi:2018oca}. 
This means that the main computational cost is in the evaluation of the decay map, which only needs to be done once, and then the final decay spectrum can be computed from an arbitrary initial particle spectrum according to \Eq{eq:map}. In particular,  more finer details of resonance decay processes could be thus efficiently treated. The primary example is a finite width of resonances, which can be included in the decay map as an additional integral over resonance mass  with, for example, Breit-Wigner distribution~\cite{Torrieri:2004zz}.  
The formalism may also be generalized to anisotropic as well as spin-dependent decays. However, even ignoring these additional complications, the sheer number of primary resonances in heavy ion collisions makes the numerical evaluation  of \Eq{eq:map} a burden. Therefore we will now specialize to the decays of initial resonance spectrum specified by a common freeze-out procedure and will leave the inclusions of resonance widths and other improvements of the decay map for future work.

\section{Cooper-Frye for the final decay spectrum\label{sec:cooperfry}}

Even after  integration of the intermediate resonances in the decay map, evaluation of the total decayed particle spectrum  requires the sum over a (possibly large) number of  primary resonances and corresponding freeze-out integrals. This sum can be performed explicitly, if we make some assumptions about the initial resonance spectra.
In the hydrodynamic description of heavy ion collisions, the initial hadron spectrum is given as
 an integral over a freeze-out hypersurface $\sigma$ according to the 
Cooper\nobreakdash-Frye formula~\cite{Cooper:1974mv}\footnote{
The hypersurface element is $d\sigma_\mu=d^3x\sqrt{h}n_\mu$, where $h$ is the determinant of the induced metric on the freeze-out surface, $d^3x\sqrt{h}$ is the invariant  volume element and $n^\mu$ is a normal vector on the surface, which we take to be pointing inwards. In this work we use mostly positive metric convention.}
\begin{equation}
E_\p\frac{d N_a}{d^3 \p} = \frac{\nu_a}{(2\pi)^3}\int_\sigma f_a(-u^\nu p_\nu, T, \mu) 
p^\mu d\sigma_\mu,\label{eq:CF}
\end{equation}
where $\nu_a$ is the degeneracy factor of spin/polarization states and $f_a$ 
is  a particle distribution function which depends on local fluid temperature $T(x)$, flow velocity $u^\mu(x)$, and chemical potential $\mu(x)$.
We will discuss more general initial particle distributions arising in dissipative hydrodynamics in \Sec{sec:perturbations}.

In the calculation of the decay particle spectrum the order of surface integral and the linear map given by \Eq{eq:map} can be reversed, resulting in the formula for the final decay particle spectrum
\begin{equation}
E_\p\frac{d N_b}{d^3 \p} = \frac{\nu_b}{(2\pi)^3}\int_\sigma g^\mu_b(p, u, T, \mu) d\sigma_\mu\label{eq:CF2},
\end{equation}
where we define \emph{vector distribution function} $g^\mu$, which for the primary resonances is $g^\mu_a = f_a p^\mu$, while for the decay products it is given by
\begin{equation}
g^\mu_b(p, u)\equiv\sum_a\frac{\nu_a}{\nu_b}\ \int\! \frac{d^3 q}{(2\pi)^3 2 E_\q}  D^a_b(p^\nu q_\nu)f_a(-u^\sigma q_\sigma) q^\mu.\label{eq:gmu}
\end{equation}
Once the function $g^\mu_b(p, u, T, \mu)$ is calculated and stored, the final decay spectra can be straightforwardly obtained by Cooper-Frye integral, \Eq{eq:CF2}, without the need of ever calculating distribution of intermediate hadrons.

If the initial distribution $f_a$ is only a function of particle energy $\bar E_\q=-q^\mu u_\mu$ in the reference frame moving with velocity $u^\mu$\footnote{In principle $u^\mu$ could be any time-like vector, not necessarily associated with a fluid flow.} and some Lorentz scalars, e.g. temperature $T$ or chemical potentials $\mu$,
then by Lorentz invariance of the decay process, the vector distribution function before and after the decay  integral in \Eq{eq:gmu} can be uniquely written as a sum of two scalar functions
\begin{equation}
g^\mu_b(p,u) = f^{\text{eq}}_{1,b}(\bar E_\p)\left(p^\mu-\bar E_\p u^\mu\right) + f^\text{eq}_{2,b}(\bar E_\p) \bar E_\p u^\mu. \label{eq:idealgmu}
\end{equation}
Here $p^\mu$, and $\bar E_\p u^\mu$ are the only available Lorentz vectors, and $f^\text{eq}_1$ and $f^\text{eq}_2$ are functions only depending on the Lorentz scalar $\bar E_\p$, and (implicitly)  $\mu$, $T$, and decay parameters.
In the fluid-restframe, 4-vectors $\bar E_\p u^\mu=(\bar E_\p, \mathbf{0})$  and $p^\mu-\bar E_\p u^\mu=(0, \bar \p)$ are two irreducible SO(3) representations transforming under rotations as a scalar and a vector, respectively. The decay operator in \Eq{eq:gmu} is a linear map and therefore guarantees that $f^\text{eq}_1$ and $f^\text{eq}_2$ components do not mix during (isotropic) decays.
The initial hadrons on the freeze-out surface are initialized by
 $g^\mu_a = f_a p^\mu $ and for the equilibrium distribution function both components $f_1^\text{eq}$ and $f_2^\text{eq}$ are initialized to be either Bose-Einstein or Fermi-Dirac distributions 
\begin{equation}
f_\text{eq}(- u^\mu p_\mu, T, \mu) = \left(e^{-u_\mu p^\mu/T-\mu/T}\mp 1\right)^{-1},
\end{equation}
where $\mu=\sum_Q \mu_Q Q$ represents the sum over the product of all relevant chemical potentials and corresponding charges.

Instead of applying the full decay map $D_b^a(q^\mu p_\mu)$ in \Eq{eq:gmu} and calculating immediately the final vector distribution function $g^\mu_b$  (from which its components $f^\text{eq}_{i,b}$ can be determined), one can also apply repeatedly the elementary 2-body and 3-body decay maps.
 This procedure is simple, because for the isotropic 2-body decay $a\rightarrow b+c$ in \Eq{eq:map2body}, the transformation rule between the parent and child  components $f_i^a$ and $f_i^b$ is simply a one dimensional integral. Leaving the details for \app{sec:decomp}, the iterative relation between the components is (c.f. \Eq{eq:mapmap})
\begin{subequations}
\begin{align}
f_{1,b}^\text{eq}(\bar E_\p) &= B \frac{\nu_a}{\nu_b} \frac{m_a^2}{m_b^2}\frac{1}{2}\int_{-1}^1 dw f_{1,a}^\text{eq}\left(E(w)\right) \frac{Q(w)}{|\bar \p|},\\
f_{2,b}^\text{eq}(\bar E_\p) &= B \frac{\nu_a}{\nu_b} \frac{m_a^2}{m_b^2}\frac{1}{2}\int_{-1}^1\! dw\, f_{2,a}^\text{eq}\left( E(w)\right) \frac{E(w)}{\bar E_\p},
\end{align}\label{eq:comp}
\end{subequations}
where we used the abbreviations
\begin{subequations}
\begin{align}
E(w) &\equiv \frac{m_a E^a_{b|c}\bar E_\p}{m_b^2}-w \frac{m_a p^a_{b|c}|\bar \p|}{m_b^2},\label{eq:Ew}\\
Q(w)&=
\frac{m_a E^a_{b|c}|\bar \p|}{m_b^2}-w \frac{m_a p^a_{b|c}\bar E_\p}{m_b^2}\label{eq:Qw},
\end{align}
\end{subequations}
and
$\bar E_\p$ and $\bar \p$ are the  energy and momentum of particle $b$ in  the fluid-restframe. 
The isotropic three body decays $a\rightarrow b+c+d$ can be easily incorporated by integrating the 2-body transformation rules \Eq{eq:comp}  over the effective decay partner mass $m_{\tilde c}$ as in \Eq{eq:3body}. Such one-dimensional integrals can be easily done by standard numerical integration routines~\cite{FastReso}.

\begin{figure}
\centering
\includegraphics[width=0.8\linewidth]{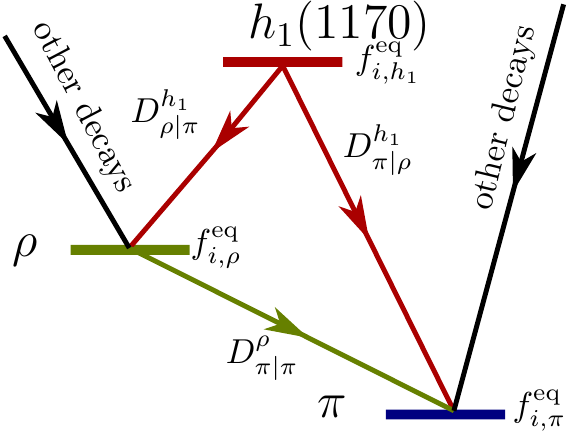}
\caption{Decay cascade $h_1\rightarrow \rho \pi \rightarrow \pi\pi\pi$. Here $h_1$ meson has only  the initial thermal distribution, while $\rho$ and $\pi$ receive feed-down from resonances' decays.}
\label{fig:decay}
\end{figure}

 For concreteness consider the decay of $h_1$ mesons illustrated in \Fig{fig:decay}. The initial irreducible components $f^\text{eq}_{i,h_1}$ for $h_1$ meson are initialized by the Bose-Einstein distribution depending on temperature and chemical potential
\begin{align}
f^\text{eq}_{i,h_1}&=f_{\text{eq},h_1}(\bar E_\p, T,\mu).
\end{align}
Then the two body decay $h_1\rightarrow \rho \pi$ produces contributions to $\rho$ and $\pi$ mesons, $f^{h_1}_{i,\rho}$ and  $f^{h_1}_{i,\pi}$, according to \Eq{eq:comp}. These have to be added to the corresponding thermal distributions and the feed-down from other resonances.
\begin{align}
f_{i,\rho}^\text{eq} &= f_{\text{eq},\rho}(\bar E_\p, T,\mu) + f^{h_1}_{i,\rho}+f^\text{other}_{i,\rho},\\
f_{i,\pi}^\text{eq} &= f_{\text{eq},\pi}(\bar E_\p, T,\mu) + f^{h_1}_{i,\pi}+f^{\rho}_{i,\pi}+f^\text{other}_{i,\pi}.
\end{align}
Here $f^{\rho}_{i,\pi}$ represents the total feed-down from $\rho\rightarrow \pi\pi$ decay irrespective of $\rho$'s origin and which is calculated according to  \Eq{eq:comp} from parent particle distribution $f_{i,\rho}^\text{eq}$. By starting from the heaviest resonance and summing the thermal and decay contributions of lower mass resonances, the irreducible components of the final stable particles can be calculated with the minimal number of  decay integrals.

\begin{figure}
\centering
\includegraphics[width=\linewidth]{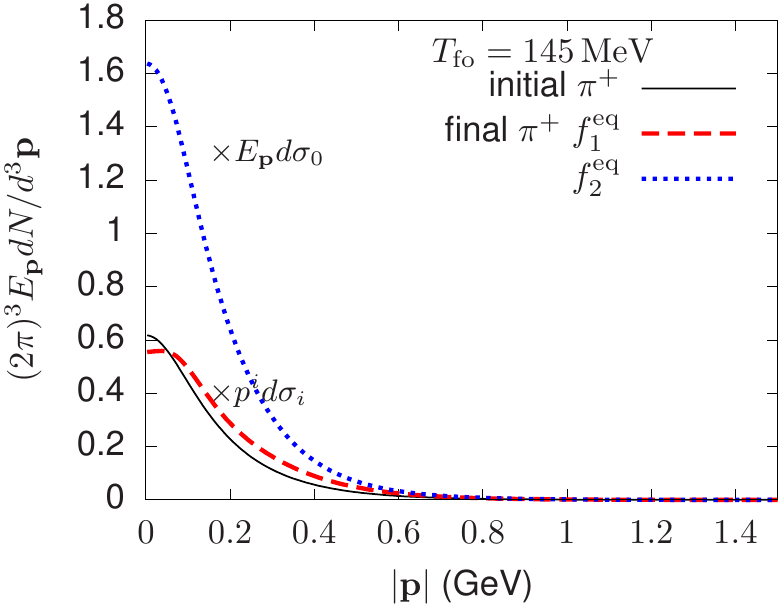}
\caption{ Lorentz invariant  pion $\pi^+$ weight functions $f_i^\text{eq}(-u^\mu p_\mu)$, \Eq{eq:idealgmu}, which determine the final pion spectrum at each space-time point on the freeze-out surface, \Eq{eq:restN}. 
Direct decays  of  $\sim 300$ hadron resonances were computed from equilibrium distributions with temperature $T_\text{fo}=145\,\text{MeV}$ and zero chemical potential $\mu_Q=0$$^\text{\ref{foot}}$. 
 The explicit prefactors indicate the required freeze-out surface elements for the Cooper-Frye integration in the fluid-restframe. Thermal pion distribution function shown for comparison.
}
\label{fig:plotfpq}
\end{figure}

The physical meaning of the irreducible components  $f^\text{eq}_{i,b}$ of the vector distribution function $g^\mu_b$ can be clarified by considering  particle spectrum in the fluid-restframe
\begin{align}
E_\p\frac{d N_b}{d^3 \p} &= \frac{\nu_b}{(2\pi)^3}\int_\sigma\left[ f^{\text{eq}}_{1,b}\left(p^\mu-\bar E_\p u^\mu\right) + f^\text{eq}_{2,b} \bar E_\p u^\mu \right] d\sigma_\mu\nonumber\\
&=\left. \frac{\nu_b}{(2\pi)^3}\int_\sigma ( f^\text{eq}_{1,b} p^i d\sigma_i +f^\text{eq}_{2,b} E_\p d\sigma_0)\right|_{u^\mu=(1,\mathbf{0})}\label{eq:restN},
\end{align}
where now $f^\text{eq}_{1,b}$ is the part of particle spectrum proportional to   $p^i d\sigma_i$ element of the freeze-out surface, while $f^\text{eq}_{2,b}$ contributes to the spectrum with $E_\p d\sigma_0$ weight.  
In \Fig{fig:plotfpq} we show the irreducible components $f^\text{eq}_1$ and $f^\text{eq}_2$ for the final pion $\pi^+$ spectrum from a completely decayed thermal $T_\text{fo}=145\,\text{MeV}$ distributions of hadron resonances as used in Monte-Carlo decay chain generators~\cite{Chojnacki:2011hb}\footnote{For simplicity of comparison, we used the default list of decay chains included in \texttt{THERMINATOR 2} package~\cite{Chojnacki:2011hb}, which includes strong and weak decays of hadron resonances with mass $<2.5\text{GeV}$\label{foot}.}. 
We see that the $f^\text{eq}_2$ component, which gives the sole contribution to the particle spectra for time-like (fixed time) freeze-out surface, is larger than thermal pion distribution due to feed-down from resonance decays, while the space-like component $f^\text{eq}_2$ remains of the same size.
In an arbitrary reference frame the decay pion spectrum can be straightforwardly calculated using frame independent formulas \Eq{eq:idealgmu} and \Eq{eq:CF2}. We will discuss this further in \Sec{sec:modebymode}.

\begin{figure}
\centering
\includegraphics[width=\linewidth]{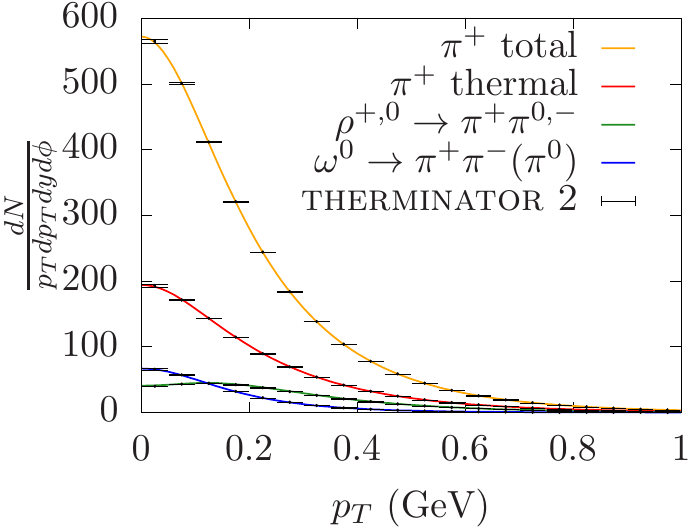}
\caption{The decay pion $\pi^+$ spectra for a simple freeze-out surface$^\text{\ref{foot2}}$ for different decay channels calculated using irreducible weight functions (see \Fig{fig:plotfpq}). Results from a Monte-Carlo generator are shown for comparison~\cite{Chojnacki:2011hb}.}
\label{fig:plotpionspectra}
\end{figure}

In \Fig{fig:plotpionspectra} we plot the final pion spectra $\pi^+$ for a simple freeze-out surface with a constant Bjorken time, freeze-out temperature and radial velocity\footnote{\label{foot2}We
used the  following \texttt{THERMINATOR 2} options  for the freeze-out surface: $\tau_\text{fo}=8.17\,\text{fm}$, $T_\text{fo}=145\,\text{MeV}$, radius of the surface  $R=8.21\,\text{fm}$  and a constant radial velocity $v_T=0.341$.}.
In addition to the total pion spectrum (which includes all decay chains producing $\pi^+$), we also show the pion spectrum from the dominant decay channels $\rho^{+,-}\rightarrow \pi^++\pi^{0,-}$ and  $\omega^0\rightarrow \pi^++\pi^-+\pi^0$ (where $\rho^{+,0}$ and $\omega^0$ spectra themselves include decay contributions from yet heavier resonances).
We compare our results with  the decay pion spectrum generated by a Monte-Carlo resonance decay generator \texttt{THERMINATOR 2}~\cite{Chojnacki:2011hb}.
 All spectra are in excellent agreement, however we would like to stress that the decay pion spectrum in  \Fig{fig:plotpionspectra} is obtained immediately from a simple Cooper-Frye freeze-out procedure \Eq{eq:CF2}. The  vector distribution function components $f_i^\text{eq}$ shown in \Fig{fig:plotfpq} only need to calculated once for a particular freeze-out temperature $T_\text{fo}$ and then can applied to any shape of the freeze-out surface or the fluid velocity field $u^\mu(x)$, without the need of costly calculations of intermediate resonances.

Although we gave an example of a constant temperature and chemical potential freeze-out surface, 
 our method  can be equally well applied for varying freeze-out temperature or chemical potential. In such case 
irreducible components of the vector distribution function $f_i^\text{eq}$ will need to be tabulated not only in the fluid-restframe energy $\bar E_\p =-u^\mu p_\mu$, but also the additional freeze-out variables. However, since this tabulation needs only to be done once, the freeze-out integral \Eq{eq:CF2} can be performed essentially without additional computational cost.

\section{Viscous and linear corrections to particle spectrum\label{sec:perturbations}}

\begin{figure*}
\centering
\includegraphics[width=0.45\linewidth]{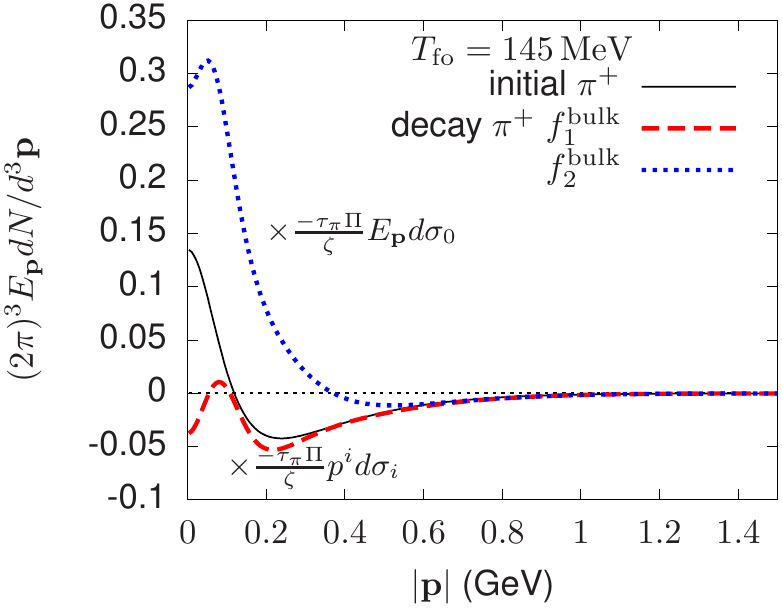}
\includegraphics[width=0.45\linewidth]{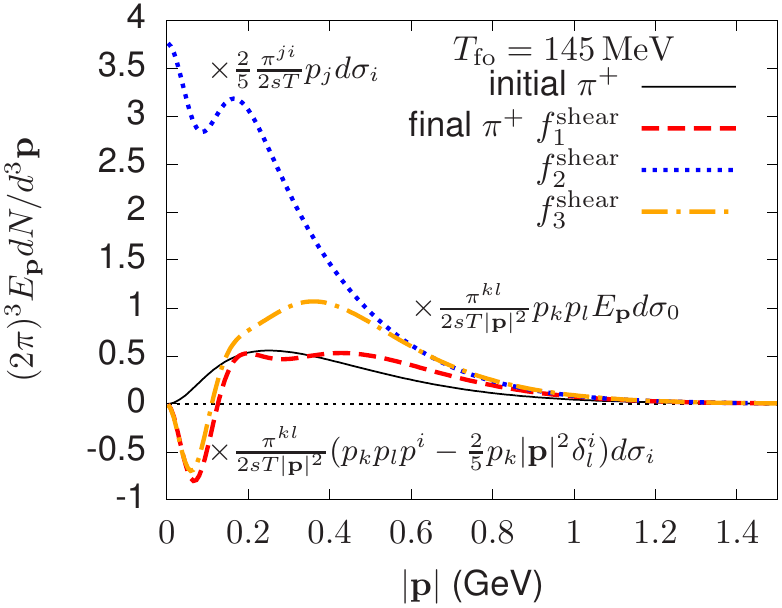}
\caption{
Lorentz invariant  pion $\pi^+$ weight functions $f_i^\text{bulk}(-u^\mu p_\mu)$ and $f_i^\text{shear}(-u^\mu p_\mu)$, \Eqs{eq:Pi} and \eq{eq:Pi}, which determine the final pion spectrum from bulk  and shear viscous perturbations of the distribution functions of primary resonances, \Eq{eq:bulkini} and \eq{eq:shearini}.
 The explicit prefactors indicate the required freeze-out surface elements for the Cooper-Frye integration in the fluid-restframe. Note that a term  $T^2/|\p|^2$ was factored out from $f_i^\text{shear}$ components.
Bulk/shear perturbation of (initial) pion distribution function shown for comparison. 
}
\label{fig:paperfjshear}
\end{figure*}

In viscous hydrodynamics, the freeze-out distribution function differs from the equilibrium Bose-Einstein or Fermi-Dirac distribution with additional dependences on the dissipative terms like the shear-stress tensor $\pi^{\mu\nu}(x)$ and the bulk-viscous pressure $\Pi(x)$, so that the initial vector distribution function in the Cooper-Frye formula, \Eq{eq:CF}, is 
\begin{equation}
g^\mu(p,u,\pi, \Pi)= f(u^\mu(x), \pi^{\rho\sigma}(x) , \Pi(x), p^\mu)p^\mu\label{eq:fgen}.
\end{equation}
The functional dependence of the distribution function on viscous corrections for hadron resonance gas is largely unresolved problem even at linear order in the dissipative terms and various parametrizations are used~\cite{Teaney:2003kp,Paquet:2015lta}. Therefore below we also consider only linear dissipative corrections to the decay particle spectrum, but note that higher order terms, if known, could be also straightforwardly included.
For concreteness we consider the form of viscous $\delta f$ corrections used in modern hydrodynamic simulations~\cite{Paquet:2015lta}
\begin{align}
&\delta f^\text{bulk}(\bar E_\p, \Pi) = f_\text{eq}(1\pm f_\text{eq})\left[\frac{\bar E_p}{T}\left(\frac{1}{3}-c_s^2\right)-\frac{1}{3}\frac{m^2}{T\bar E_p} \right]\frac{\tau_\Pi \Pi}{\zeta},\label{eq:bulkini}\\
&\delta f^\text{shear}(\bar E_\p, \pi_{\rho\nu}p^\rho p^\nu) = f_\text{eq}(1\pm f_\text{eq})\frac{\pi_{\rho\nu}p^\rho p^\nu}{2(e+p)T^2}.\label{eq:shearini}
\end{align}
Here $c_s(T)$ is the speed of sound of the medium at the freeze-out temperature, $m$ -- mass of the primary resonance, and $\tau_\Pi/\zeta$ is the ratio of bulk relaxation time and bulk viscosity.

We want to compute the final decay particle spectrum arising from such viscous components. The procedure is analogous to the one described in the previous section. First we expand \Eq{eq:fgen} to linear order in viscous corrections
\begin{equation}
g^\mu = g^\mu_\text{eq}+  g_\text{shear}^{\mu \rho \sigma}\pi_{\rho\sigma}+g^\mu_\text{bulk} \Pi+\ldots
\end{equation} 
where the derivatives $g_\text{shear}^{\mu \rho \sigma}\equiv \partial g^{\mu}/\partial {\pi_{\rho\sigma}}$ and $g_\text{bulk}^\mu\equiv \partial g^{\mu}/\partial \Pi $ can only be functions of 4-vectors $p^\mu$ and $\bar E_\p u^\mu$, and Lorentz scalars like temperature, chemical potential or resonance mass.
Initially they are given by \Eq{eq:bulkini} and \eq{eq:shearini}
\begin{equation}
 g^\mu_\text{bulk} \Pi= \delta f^\text{bulk} p ^\mu, \quad  g_\text{shear}^{\mu \rho \sigma}\pi_{\rho\sigma}= \delta f^\text{shear}p^\mu.
\end{equation}
After the decays $g^\mu_\text{bulk}$ and $g_\text{shear}^{\mu \rho \sigma}$ can be written uniquely as a certain sum of Lorentz vectors/tensors.
In \app{sec:decomp} we discuss the general irreducible decomposition of Lorentz tensors in terms of  representations transforming differently under SO(3) rotations in the fluid-restframe. Here we only reproduce the final result for the bulk
\begin{align}
g^{\mu}_\text{bulk}\Pi  =\Big[\left( p^\mu-\bar E_\p u^\mu\right)&f^\text{bulk}_1(\bar E_\p)\nonumber\\
+  \bar E_\p u^\mu &f^\text{bulk}_2(\bar E_\p)\Big]\times \frac{-\tau_\pi\Pi}{\zeta}\label{eq:Pi},
\end{align}
and shear perturbations
\begin{align}
g^{\mu\nu\rho}_\text{shear} \pi_{\nu\rho} = \Big\{& [\eta^{\rho \sigma}(p^\mu-\bar E_\p u^\mu)-\frac{2}{5} \eta^{\rho\mu} \Delta^{\sigma \alpha}p_\alpha] f^\text{shear}_1(\bar E_\p)\nonumber\\
&+ \frac{2}{5}\eta^{\rho\mu} \Delta^{\sigma \alpha}p_\alpha f^\text{shear}_2(\bar E_\p)\nonumber\\
&+ \eta^{\rho \sigma} \bar E_\p u^\mu f^\text{shear}_3(\bar E_\p)\Big\}\times \frac{p^\nu \pi_{\nu\rho}p_\sigma}{2(e+p)T^2}.\label{eq:pi}
\end{align}
The bulk pressure perturbation does not introduce new tensor structures and the decomposition is the same as for the equilibrium distribution in \Eq{eq:idealgmu}, but the initial distribution functions $f^\text{bulk}_i$  are, of course, different and can be read off from \Eq{eq:bulkini}. The linear perturbations in the shear-stress tensor induces a rank-3 tensor distribution function  $g^{\mu\nu\rho}_{\text{shear}}$, which has three non-vanishing irreducible components $f^\text{shear}_1$, $f^\text{shear}_2$ and $f^\text{shear}_3$ corresponding to a symmetric traceless tensor, vector and scalar representations (see \app{sec:decomp}).
The irreducible weight functions $f_i$ of final decay particle distribution can be calculated iteratively using similar integrals as for the equilibrium distribution \Eq{eq:comp}
\begin{align}
f_i^b(\bar E_\p) &= B \frac{\nu_a}{\nu_b}\frac{m_a^2}{m_b^2 }\frac{1}{2}\int_{-1}^1 dw A_i(w) f_i^a(E(w)).
\end{align}
where integration measures $A_i(w)$ are listed in \app{sec:decomp}.

 In \Fig{fig:paperfjshear} we show the final decayed pion $\pi^+$ spectrum in the fluid-restframe  due to viscous perturbations at $T_\text{fo}=145\,\text{MeV}$\footnote{At $T_\text{fo}=145\,\text{MeV}$ the sound velocity needed for bulk perturbation \Eq{eq:bulkini} is $c_s^2(T_\text{fo})\approx 0.14$~\cite{Borsanyi:2016ksw}.}. Different lines in \Fig{fig:paperfjshear}  correspond to different contributions stemming from components $f_i$ in \Eq{eq:Pi} and \Eq{eq:pi}. The labels next to the lines indicate the required factors for the Cooper-Frye freeze-out integral in the fluid-restframe. Note that we factored in $|\bar \p|^2$ for the shear perturbations.
 We also factored out the terms proportional to the transport coefficients, so  any (small) viscous perturbation will produce the same correction to the particle spectrum (up to the magnitude) in the local fluid-restframe. 
However the presence of such viscous corrections in the hydrodynamic evolution modifies the fluid velocity and temperature fields, and therefore the freeze-out surface will itself be different. Then evaluating the generalized Cooper-Frye freeze-out integral \Eq{eq:CF2} will yield different particle spectrum.

\begin{figure*}
\centering
\includegraphics[width=0.45\linewidth]{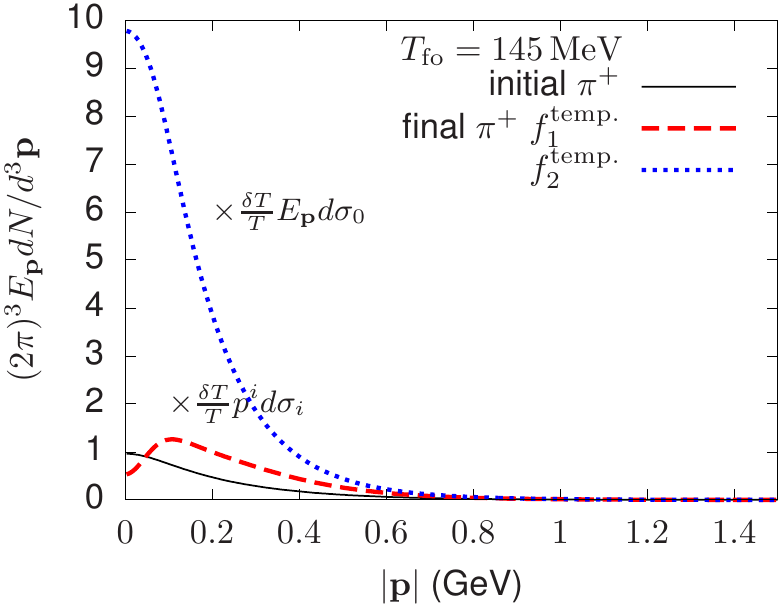}
\includegraphics[width=0.45\linewidth]{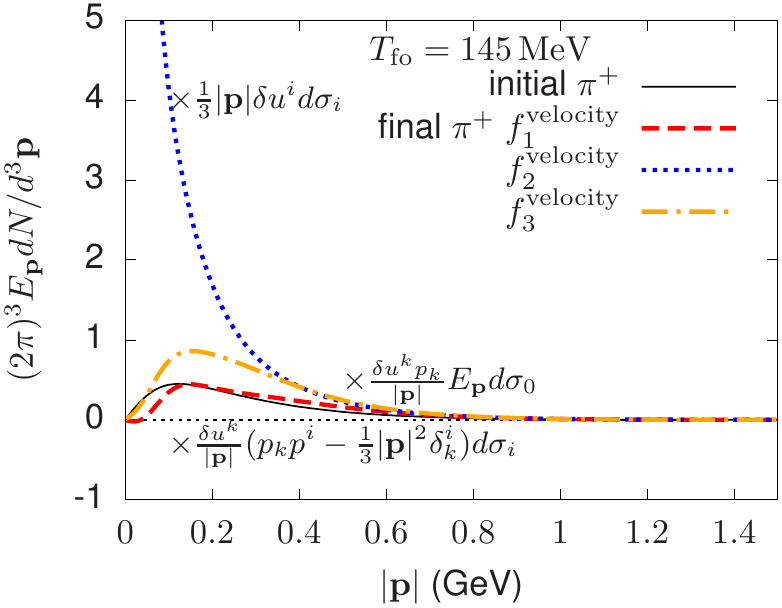}
\caption{
Lorentz invariant  pion $\pi^+$ weight functions $f_i^\text{temp.}(-u^\mu p_\mu)$ and $f_i^\text{velocity}(-u^\mu p_\mu)$, \Eqs{eq:Pi} and \eq{eq:Pi}, which determine the final pion spectrum from temperature  and velocity perturbations of the distribution functions of primary resonances, \Eq{eq:deltaT} and \eq{eq:deltau}.
 The explicit prefactors indicate the required freeze-out surface elements for the Cooper-Frye integration in the fluid-restframe.   Note that a term  $\bar T/|\p|$ was factored out from $f_i^\text{velocity}$ components.
Temperature/velocity perturbation of (initial) pion distribution function shown for comparison. }
\label{fig:paperfjscalar}
\end{figure*}

Similarly to viscous perturbations, one can also consider linear perturbations of fluid velocity $\delta u^\mu$, or temperature $\delta T$ around the background fields $\bar u^\mu$ and $\bar T$\footnote{Note that a constant temperature freeze-out surface depends on $\delta T$. However, one can also consider a freeze-out at a constant \emph{background} temperature $\bar T$, which is then independent of the perturbation.},
\begin{align}
\delta f^\text{temp.}(\bar E_\p, \delta T) &= f_\text{eq}(1\pm f_\text{eq})\frac{\bar E_\p }{\bar T}\frac{\delta T}{\bar T},\\
\delta f^\text{velocity}(\bar E_\p, \delta u_\mu p^\mu) &= f_\text{eq}(1\pm f_\text{eq})\frac{\delta u_\nu p^\nu}{\bar T}.
\label{eq:deltaflinearperturbations}
\end{align}
The irreducible decomposition for temperature perturbation is the same as for the equilibrium distribution
\begin{align}
g^{\mu}_\text{temp.} \delta T = \Big[ &\left(p^\mu-\bar E_\p u^\mu\right)f_1^\text{temp.}(\bar E_\p) \nonumber\\
&+  \bar E_\p u^\mu f_2^\text{temp.}(\bar E_\p)\Big]\times \frac{\delta T}{\bar T} \label{eq:deltaT}.
\end{align}
The irreducible  decomposition for velocity perturbations consists of symmetric traceless tensor, vector and scalar representations with corresponding weight functions $f_i^\text{velocity}$
\begin{align}
g^{\mu\nu}_\text{vel.}\delta u_\nu=\Big\{&[\eta^{\nu\rho}(p^\mu-\bar E_\p u^\mu)-\frac{1}{3} \eta^{\nu\mu}p_\sigma \Delta^{\sigma \rho} )]f_1^\text{velocity}(\bar E_\p)\nonumber\\
&+\frac{1}{3} \eta^{\nu\mu}p_\sigma \Delta^{\sigma \rho} f_2^\text{velocity}(\bar E_\p)\nonumber\\
&+ \eta^{\nu\rho}\bar E_\p u^\mu f_3^\text{velocity}(\bar E_\p)\Big\}\times\frac{\delta u_{\nu} p_\rho}{\bar T}\label{eq:deltau}.
\end{align}
The total vector distribution function is then given by
\begin{equation}
g^\mu_\text{ideal} = \bar g^\mu+  g_\text{vel.}^{\mu \rho}\delta u_{\rho}+g^\mu_\text{temp.} \delta T+\ldots
\label{eq:gmuFluidPerturbations}
\end{equation}
In \Fig{fig:paperfjscalar} we show the particle spectrum decomposition due to temperature and velocity perturbation for the same freeze-out temperature. As before we factor out the explicit dependence on the magnitude of the perturbation in the fluid-restframe.
Such linearised perturbations can be used to study, for example, the angular velocity modulations around a known freeze-out flow $u^\mu$, e.g. provided by a blast-wave model or in the  mode by mode description of heavy ion collisions~\cite{Floerchinger:2013rya,Floerchinger:2013hza}.

We would like to note in passing that other types of perturbations to the equilibrium spectrum can be considered. At lower collision energies, as in the Beam Energy Scan studies, the freeze-out distribution function will also depend on the chemical potential $\mu_Q(x)$ and one could consider linear perturbations of the chemical potential $\delta \mu_Q$, which are treated identically to the temperature variations \Eq{eq:deltaT}. Also, the hydrodynamics at non-zero baryon density must also evolve the baryon current $j^\mu=n_B u^\mu +j^\mu_D$ (e.g.\ see \cite{Floerchinger:2015efa}). The transverse (diffusion) part of this current will induce perturbations of the freeze-out surface distribution functions analogous to velocity perturbations in \Eq{eq:deltau}
\begin{equation}
\delta f^\text{diffusion}(\bar E_\p, j^\mu_D p_\mu)=f_\text{eq}(1\pm f_\text{eq})\left[\frac{n_B}{e+p}-\frac{Q_B}{\bar E_\p}\right]\frac{j_D^\nu p_\nu}{\hat \kappa}.\label{eq:diffusion}
\end{equation}
where $n_B$ is the local baryon density, $Q_B$ is the baryon charge and  $\hat \kappa_B$ is a related transport coefficient~\cite{Denicol:2018wdp}. Such additional terms are easy to accommodate in our framework, because the same transformation rules can be used to find the final decay spectrum (see \app{sec:decomp}).

\section{Application to blast-wave and mode-by-mode freeze-out\label{sec:modebymode}}

In this section we will discuss practical implementation of the fast resonance decay procedure in heavy ion collisions. Collecting together all equilibrium and viscous terms contributing to the particle spectrum, we can group them by how they are contracted with the freeze-out surface element $d\sigma_\mu$
\begin{equation}
 E_\p\frac{d N}{d^3 \p} =   \frac{\nu}{(2\pi)^3} \int_\sigma d\sigma_\mu {\Big \{} F p^\mu  + G  \bar E_\p u^\mu  + H \frac{p^\nu \pi_\nu^{\;\;\mu}|\bar \p|^2}{2(e+p)T^2} {\Big \}},
\label{eq:fototal}
\end{equation}
where explicitly these terms are
\begin{align}
F=&f^\text{eq}_1(\bar E_\p) + f_1^\text{shear}(\bar E_\p) \frac{\pi_{\rho\sigma}p^\rho p^\sigma}{2(e+p)T^2} +  f_1^\text{bulk}(\bar E_\p) \frac{-\tau_\pi\Pi}{\zeta} ,\nonumber\\
G=&f^\text{eq}_2(\bar E_\p)-f^{\text{eq}}_1(\bar E_\p) 
+ \left(f_2^\text{bulk}(\bar E_\p) - f_1^\text{bulk}(\bar E_\p) \right) \frac{-\tau_\pi\Pi}{\zeta}\nonumber\\&
 + \left(f_3^\text{shear}(\bar E_\p) - f_1^\text{shear}(\bar E_\p)\right) \frac{\pi_{\rho\sigma}p^\rho p^\sigma}{2(e+p)T^2} ,\nonumber\\
H=& \left(f_2^\text{shear}(\bar E_\p) -f_1^\text{shear}(\bar E_\p) \right) \frac{2}{5}  .
\label{eq:backgroundSpectrum1}
\end{align}
The required values of fluid velocity $u^\mu$, shear and bulk stresses $\pi^{\mu\nu}$, $\Pi$ on the freeze-out surface must be provided by the hydrodynamic model of the QGP fireball or freeze-out parametrization, e.g. blast-wave model~\cite{Schnedermann:1993ws}.  Then the complete final decay particle spectrum (for a particle species or even for the total sum of charged particles) can be computed according to \Eq{eq:fototal} by using the tabulated values of irreducible weight functions $f_i$.

Up to now the freeze-out surface was left completely general.
An interesting application of our formalism arises from the mode-by-mode solution to the fluid dynamic expansion of a fireball~\cite{Floerchinger:2013rya,Floerchinger:2013hza}. In that formalism, one decomposes the fluid fields (temperature, chemical potentials, fluid velocity, shear stress and bulk viscous pressure) into a background part and a fluctuating part, e.g.\ $u^\mu = \bar u^\mu+\delta u^\mu$. 
 In high energy collisions, a boost invariance is a good symmetry and the collision is often parametrized in Bjorken coordinates
\begin{equation}
ds^2=-d\tau^2 + dr^2 + r^2 d\phi^2 + \tau^2 d\eta^2.\label{eq:coord}
\end{equation}
It is particularly convenient to take the background part as symmetric with respect to azimuthal rotations and longitudinal Bjorken boosts (see e.g.\ \cite{Floerchinger:2017cii} for the fluid equations of motion in this situation).
Then the hadron spectrum resulting from the fluid fields after freeze-out can be also split into a background, which is invariant under these symmetries (now in momentum space), and a fluctuating part.
The freeze-out surface in this case is given by a 1D curve in $\tau\text{--}r$ plane, which can be conveniently parametrized by $(\tau(\alpha), r(\alpha) )$ where $\alpha\in (0,1)$ and
\begin{equation}
d\sigma_\mu =  \tau(\alpha) r(\alpha) \left( \frac{\partial r}{\partial \alpha}, -\frac{\partial \tau}{\partial \alpha}, 0, 0 \right) d\alpha d\phi d\eta
\label{eq:hypersurfaceelement}
\end{equation}
Similarly by symmetry the fluid velocity has only two components and can be written in terms of a radial fluid rapidity $\bar \chi$, 
\begin{equation}
\bar u^\mu = \left( \cosh(\bar \chi), \sinh(\bar \chi), 0, 0 \right).\label{eq:velocity}
\end{equation}
Note that then the particle energy in fluid-restframe is
\begin{equation}
\bar E_\p = m_T \cosh(\bar \chi) \cosh(\eta-y) - p_T \sinh(\bar \chi) \cos(\phi-\varphi).\label{eq:ebar}
\end{equation}
where in the coordinate system of \Eq{eq:coord} and at space time point $(\tau,r,\phi,\eta)$ the particle momentum components are given by
\begin{equation}
\begin{split}
p^{\tau}=m_T \cosh(\eta-y),\quad p^{r}=p_T \cos(\phi-\varphi), \\
p^{\phi}=\frac{p_T}{r} \sin(\phi- \varphi),\quad p^{\eta}=\frac{m_T}{\tau} \sinh(\eta-y).
\label{eq:momentumBjorkenCoord}
\end{split}
\end{equation}
Here we use the transverse momentum $p_T$ (and transverse mass $m_T=\sqrt{m^2+p_T^2}$), the particle momentum angle $\varphi$, 
and the particle momentum rapidity $y$.

The background contribution to the shear stress tensor can be parametrized in terms of two independent components, here taken to be $\bar \pi^\phi_\phi$ and $\bar \pi^\eta_\eta$
\begin{equation}
\begin{split}
& \bar \pi_{\tau\tau} = -\bar u^r \bar  u^r\left[ \bar \pi^\phi_\phi + \bar \pi^\eta_\eta \right], \quad \bar \pi_{\tau r } = \bar \pi_{r\tau } =\bar  u^r \bar  u^\tau \left[ \bar \pi^\phi_\phi + \bar \pi^\eta_\eta \right], \\
& \bar\pi_{rr} = -\bar u^\tau \bar  u^\tau  \left[ \bar \pi^\phi_\phi + \bar \pi^\eta_\eta \right],
\quad  \bar \pi_{\phi\phi} = r^2 \bar \pi^\phi_\phi , \quad \bar\pi_{\eta \eta} = \tau^2 \bar\pi^\eta_\eta,
\end{split}
\label{eq:backgroundshearstress}
\end{equation}
and the bulk viscous pressure is simply $\bar \Pi$.

The decay hadron spectrum for azimuthally and boost invariant freeze-out surface then reduces to a single integral 
\begin{equation}
\begin{split}
\frac{d N}{2\pi p_Tdp_Tdy} &= \frac{\nu}{(2\pi)^3} \int_0^1\! d\alpha \; \tau(\alpha) r(\alpha)  
\\
 {\Bigg \{} \frac{\partial r}{\partial\alpha}
\Big[
K^\text{eq}_1
&+ \frac{\bar \pi^\eta_\eta}{2(e+p)T^2} \; K_1^\text{shear}
+ \frac{\bar \pi^\phi_\phi}{2(e+p)T^2}  \; K_3^\text{shear} \\
&+ \frac{-\tau_\pi \bar \Pi}{\zeta}\; K_1^\text{bulk}
\Big]
\\
-\frac{\partial \tau}{\partial\alpha}
\Big[
K^\text{eq}_2 &+ \frac{\bar \pi^\eta_\eta}{2(e+p)T^2}  \; K_2^\text{shear}+ \frac{\bar\pi^\phi_\phi}{2(e+p)T^2}  \; K_4^\text{shear} \\
&+  \frac{-\tau_\pi\bar \Pi}{\zeta} \; K_2^\text{bulk}
\Big]
{\Bigg \}},
\end{split}
\label{eq:backgroundspectrumintermsofkernels}
\end{equation}
were the freeze-out kernels $K_i(p_T, \bar \chi)$ are solely functions  of the transverse particle momentum and radial fluid velocity $\bar u^r=\sinh \bar \chi$, and the terms proportional to viscous tensors are factored out. Analogously to the original irreducible weights $f_i$,  the freeze-out kernels can be precomputed and applied to an arbitrary freeze-out surface ($\tau(\alpha), r(\alpha)$) and radial fluid rapidity profile $\bar \chi(\alpha)$.
For the equilibrium components $f_i^\text{eq}$ these kernels are given by the following rapidity and angle integrals
\begin{equation}
\begin{split}
K^\text{eq}_1(p_T,\bar\chi)  =  & \int_0^{2\pi} \!d\phi \int_{-\infty}^\infty\! d\eta  \left\{ f^\text{eq}_1(\bar E_\p) m_T \cosh(\eta-y)  
\right.\\
&
\left.
+ \left(f^\text{eq}_2(\bar E_\p) - f^\text{eq}_1(\bar E_\p)\right) \bar E_\p \cosh(\bar \chi) \right\}, \\
K^\text{eq}_2(p_T,\bar\chi)  = & \int_0^{2\pi}\! d\phi \int_{-\infty}^\infty\! d\eta \left\{ f^\text{eq}_1(\bar E_\p) p_T \cos(\phi-\varphi) 
\right.\\
&
\left. + \left(f^\text{eq}_2(\bar E_\p) - f^\text{eq}_1(\bar E_\p)\right) \bar E_\p \sinh(\bar \chi) \right\}. \\
\end{split}
\end{equation}
Recall that $\bar E_\p$ also depends only on the differences of $\eta-y$ and $\phi-\varphi$, \Eq{eq:ebar}, so the dependence on momentum rapidity $y$ and angle $\varphi$ disappears after the integration. The integral expressions for viscous kernels are given in \app{sec:kernels}.
For the simple constant time freeze-out surface used in \Fig{fig:plotpionspectra} only the temporal part of the freeze-out surface contributes and the decay pion spectrum is proportional to $K_1^\text{eq}(p_T,\bar\chi=\arctan v_T)$.

All deviations from the azimuthally and boost invariant background in mode-by-mode hydrodynamics is carried by the fluctuating part of the fluid fields and to first approximation only the linear part in these perturbations contribute to the final particle spectra. 
For perturbations in e.g.\ fluid velocity $\delta u^\mu$ the corresponding perturbations of the distribution functions are given explicitly in \Eq{eq:deltaflinearperturbations}. 
Because the decay operator \Eq{eq:map} is linear, if $\delta u^\mu$ is written  as linear superposition of Fourier modes in azimuthal angle $\phi$ and space-time rapidity $\eta$
\begin{equation}
\delta u^\mu(x) = \left( \tanh(\bar \chi) \delta u^r, \delta u^r, \delta u^\phi, \delta u^\eta \right) e^{im\phi + i k \eta},
\label{eq:fluidvelocityperturbation}
\end{equation}
one finds that the distribution of hadrons after kinetic freeze-out and resonance decays depends on momentum space azimuthal angle $\varphi$ and momentum space rapidity $y$ 
via the combination $e^{im\varphi+iky}$ with the same azimuthal wavenumber $m$ and rapidity wavenumber $k$. This is a direct consequence of $\text{U}(1)\times \mathbb{R}^1$ symmetry, which prevents different representations labelled by $m$ and $k$ from mixing under linear operations. This way arbitrary linear perturbations in fluid fields can be mapped to the modes of the final particle spectra, which can be straightforwardly incorporated in the formalism of mode-by-mode hydrodynamics~\cite{Floerchinger:2013rya,Floerchinger:2013hza}.

\section{Conclusions\label{sec:conclusions}}
We presented a method to calculate the final decay spectrum of direct resonance decays \emph{directly} from hydrodynamic fields on a freeze-out surface.
By applying the decay map, \Eq{eq:map}, to the distribution function of primary particles \emph{before} the Cooper-Frye integration, we found the (vector) distribution function of decay products, \Eq{eq:gmu}.
By  decomposing this distribution into components transforming differently under SO(3) rotations in the fluid-restframe, we expressed the final decay particle spectrum as a sum of a few Lorentz invariant weight functions and known Lorentz vectors.
The explicit procedure to determine the irreducible weight functions for an arbitrary decay chain of isotropic  2-body and 3-body decays was derived and a numerical implementation  was made public~\cite{FastReso}.
  We considered primary hadron resonances generated by the equilibrium distribution function, viscous shear and bulk perturbations, and linearised temperature and velocity perturbations. Modifications to the particle spectrum due to variations in the chemical potential and the diffusive part of particle current can be also straightforwardly included in this framework.

The final 1-body particle spectrum of decay products is then calculated from a general Cooper-Frye-type freeze-out integral, \Eq{eq:fototal}.
The most important aspect of our method is that intermediated particle decays do not need to be calculated event-by-event. The irreducible components of the decay particle distribution function \Eq{eq:gmu} are computed only once, and the spectrum of a few relevant hadron species, which includes  feed-down of all direct decays, can be computed for an arbitrary freeze-out surface.
This significantly reduces the computational costs of direct resonance decays. 

Although our method of calculating direct resonance decays is already competitive with other treatments available on the market, the computational efficiency of our approach makes it practical to include finer details of resonances decays.  For example, new hadron resonance states can be easily added to improve the agreement between the lattice QCD and hadronic equation of state~\cite{Alba:2017hhe}. Finite widths of the resonances can be incorporated in the decay map~\cite{Torrieri:2004zz,Lo:2017sux,Huovinen:2016xxq}. This has recently been shown to reduce the discrepancy between measured and predicted proton yield in the statistical hadronization models~\cite{Vovchenko:2018fmh,Andronic:2018qqt}.

In this work we neglected hadronic rescatterings after the chemical freeze-out, which may change the final particle spectra, but the effect is subleading in comparison with the decay feed-down~\cite{Ryu:2017qzn}. Elastic scatterings in the hadronic phase can be modeled by a hydrodynamic evolution of hadron fluid in partial chemical equilibrium~\cite{Bebie:1991ij,Hirano:2002ds}. In this scenario, the particle ratios are fixed at the chemical freeze-out temperature $T_\text{chem}$ for each species $i$ of long lived hadrons by introducing an appropriate chemical potentials $\mu_i(T)$ for $T<T_\text{chem}$. Subsequently, the kinetic freeze-out may take place at some lower temperature $T_\text{kin}$. Because the primary resonance spectra are still described only by temperature $T_\text{kin}$ and the chemical potentials $ \mu_i(T_\text{kin})$, the direct decays can be calculated using the techniques proposed in this work.

Another interesting generalization of the framework is to keep track of the particle spin in the decays. This could be particularly useful for the studies of vorticity polarization in heavy ion collisions~\cite{Becattini:2013fla,Becattini:2016gvu}. However, in this case one might need to go beyond  isotropic $s$-wave decays and consider more general momentum dependent decay patterns, which to our knowledge were not included in phenomenological works so far. 
Finally, we note that the 1-particle distribution function does not have the information of the connected two-particle function, namely the non-flow correlations of particles produced by the same resonance decay. 
However, the decay map \Eq{eq:map}  can be generalized to two-particle spectrum.

In summary, we believe that the computationally efficient way of computing direct resonance decays, which was presented in this paper, will be of great practical utility for phenomenological studies of heavy ion collisions and make realistic particle yield calculations much more affordable.

\noindent{\bfseries Acknowledgments:}
The authors would like to thank Gabriel Denicol, Ulrich Heinz, Jacquelyn Noronha-Hostler, Jean-Fran\c cois Paquet, and Iurii Karpenko for useful discussions and comments.
This work is part of and supported by the DFG 
Collaborative Research Centre SFB 1225 (ISOQUANT) (S.F., E.G., A.M.). This work was supported in part by the U.S. Department of Energy, Office of 
Science, Office of Nuclear Physics  under Award Numbers 
DE\nobreakdash-FG02\nobreakdash-88ER40388 (D.T.).

\appendix
\section{Tensor decomposition and decay rules\label{sec:decomp}}

Instead of calculating the decay operator to a particle spectrum, \Eq{eq:map}, one may also apply elementary decay operators directly to the distribution function and obtain thus a modified distribution function, \Eq{eq:gmu}, that already includes resonance decays. A technical complication is that one needs to do this based on a Lorentz-vector form of the distribution function, which  at the freeze-out surface is given by
\begin{equation}
g^\mu = p^\mu f(p,u,T,\mu).
\end{equation}
 In ideal hydrodynamics $f$ is simply Bose-Einstein or Fermi-Dirac distribution
\begin{equation}
\label{eq:eq}f_\text{eq}(- u^\mu p_\mu, T, \mu) = \left(e^{-u_\mu p^\mu/T-\mu/T}\mp 1\right)^{-1}.
\end{equation}
 Note that although the decay operator is Lorentz invariant, the Lorentz boost symmetry is explicitly broken by the fluid velocity, which singles out a reference frame. The residual rotational subgroup allows to decompose $g^\mu$ into  two 4-vectors transforming under different representations of SO(3) in the fluid-restframe
\begin{equation}
g^\mu_\text{eq}= \hat p^\mu f_1^\text{eq} +\hat q^\mu  f_2^\text{eq},
\label{eq:vectordistributionfunction}
\end{equation}
where ($\Delta^{\mu\nu}=\eta^{\mu\nu}+u^\mu u^\nu$)
\begin{equation}
\hat q^\mu = \bar E_\p u^\mu, \quad \text{and} \quad \hat p^\mu = p_\mu \Delta^{\nu\mu}=p^\mu - \bar E_\p u^\mu,
\end{equation}
transform as a scalar ($\bf 1$) and a vector ($\bf 3$) respectively. 
Both representations have  identical weight functions $f_1^\text{eq}=f_2^\text{eq}=f_\text{eq}$ on the freeze-out surface, but after the resonance decays are taken into account the two functions will differ. {Isotropic} decays do not mix different SO(3) representations and the new weight functions are found by successively  applying the decay maps \Eq{eq:map2body} and \Eq{eq:3body}. Before deriving the specific transformation rules for the weight functions $f_i$, lets consider a more general case for the distribution function. In viscous hydrodynamics, the freeze-out distribution function differs from the equilibrium expression \Eq{eq:eq} and also depends on shear-stress tensor and bulk viscous pressure
\begin{equation}
g^\mu = f(-u_\mu p^\mu, \pi_{\rho\sigma} p^\rho p^\sigma, \Pi, T, \mu)p^\mu.
\end{equation}
Close to equilibrium one may Taylor expand the vector distribution function around the equilibrium distribution function
\begin{equation}
g^\mu = g^\mu_\text{eq}+  g_\text{shear}^{\mu \rho \sigma}\pi_{\rho\sigma}+g^\mu_\text{bulk} \Pi+\ldots\label{eq:taylor1},
\end{equation} 
where the derivatives $g_\text{shear}^{\mu \rho \sigma}\equiv \partial g^{\mu}/\partial {\pi_{\rho\sigma}}$ and $g_\text{bulk}^\mu\equiv \partial g^{\mu}/\partial \Pi $ can only be functions of 4-vectors $p^\mu$ and $\bar E_\p u^\mu$, and Lorentz scalars like temperature, chemical potential or resonance mass.
Similarly, we can also consider perturbations of the hydrodynamic fields around an arbitrary background, e.g.\ $u^\mu=\bar u^\mu +\delta u^\mu$ and $T = \bar T + \delta T$. Then the vector distribution function decomposes as
\begin{equation}
g^\mu = \bar g^\mu+  g_\text{velocity}^{\mu \rho}\delta u_{\rho}+g^\mu_\text{temp.} \delta T+\ldots\label{eq:taylor2}
\end{equation}
Thanks to the linearity of the decay map, each term in the Taylor expansion ($g^\mu, g^{\mu\rho}$, $g^{\mu\rho\sigma }$) can be also decomposed into irreducible representations of SO(3) rotational group.
For example, a two-tensor distribution function $g^{\mu\nu}(u, p)$ contains irreducible representations of $\text{SO}(3)$ according to the tensor decomposition $({\bf 1}+{\bf 3})\otimes({\bf 1}+{\bf 3}) = 2 \times {\bf 1}+ 3\times {\bf 3} + {\bf 5} $. In terms of the 4-vectors $\hat q^\mu$ and $\hat p^\mu$ they are given as
\begin{enumerate}
\item[a)] two scalars,
\begin{equation}
1)\quad \hat q^\mu \hat q^\nu \quad\quad\quad\text{and} \quad\quad 2) \quad |\bar \p|^2 \Delta^{\mu\nu},
\label{eq:4.7}
\end{equation}
\item[b)] three vectors,
\begin{equation}
1)\quad \hat q^\mu \hat p^\nu, \quad 2)\quad \hat p^\mu \hat q^\nu, \quad \text{and} \quad 3)\quad \hat q_\alpha  \hat p_\beta  \epsilon^{\alpha \beta \mu\nu},
\label{eq:4.8}
\end{equation}
\item[c)] a traceless, symmetric two-tensor,
\begin{equation}
\left(  \hat p^\mu \hat p^\nu - \tfrac{1}{3} |\bar \p|^2 \Delta^{\mu\nu} \right)
\label{eq:4.9}.
\end{equation}
\end{enumerate}
However, in practice not all of these terms are needed. Thanks to the orthogonality relation $\bar u^\rho \delta u_\rho = 0$, 
the first terms in \Eq{eq:4.7} and \Eq{eq:4.8} drop out, while the antisymmetric  term in \Eq{eq:4.8} is not present initially and by symmetry reasons is not generated by the decay operator. 
Only one scalar, one vector  and one tensor representation contribute to the contraction $g^{\mu\nu}_\text{velocity} \delta u_\nu$
\begin{align}
&g^{\mu\nu}_\text{velocity}\delta u_\nu =\Big\{[\delta u_{\nu} p^\nu(p^\mu-\bar E_\p u^\mu)-\frac{1}{3}|\bar \p|^2 \delta u^\mu)]f_1^\text{velocity}\nonumber\\
&+\frac{1}{3}|\bar \p|^2\delta u^\mu f_2^\text{velocity}+ \delta u_{\nu}p^\nu\bar E_\p u^\mu f_3^\text{velocity}\Big\}\times \frac{1}{\bar T}\label{eq:dudecomp}
\end{align}

Analogously we list the most general decomposition of the  three-tensor distribution function $g^{\mu\nu\rho}$ into the $\text{SO}(3)$ representations $({\bf 1}+{\bf 3})\otimes ({\bf 1}+{\bf 3})\otimes ({\bf 1}+{\bf 3}) = 5 \times {\bf 1}+ 9\times {\bf 3} + 5\times{\bf 5} + {\bf 7} $
\begin{enumerate}
\item[a)] five scalars
\begin{align}
&1)\quad \hat q^\mu \hat q^\nu \hat q^\rho, \quad 2)\quad  |\bar \p|^2 \hat q^\mu\Delta^{\nu\rho},  \text{ and 2 perm.} \\ &3)\quad  \hat q_\alpha \epsilon^{\alpha\mu\nu\rho} |\bar \p|^2  .
\end{align}
\item[b)] nine vectors 
\begin{align}
&1)\quad \hat q^\mu \hat q^\nu \hat p^\rho,\, \text{ and 2 perm.} \\
&2)\quad \hat q_\alpha  \hat p_\beta \epsilon^{\alpha\beta \mu\nu}  \hat q^\rho, \text{ and 2 perm.}\\
&3)\quad |\bar \p|^2 \Delta^{\mu\nu} \hat p^\rho \text{ and 2 perm.}.
\label{eq:4.11}
\end{align}
\item[c)] five symmetric and traceless two-tensors
\begin{align}
&1)\quad \hat q^\mu \left(  \hat p^\nu \hat p^\rho - \tfrac{1}{3}\Delta^{\nu\rho}|\bar \p|^2\right), \text{ and 2 perm.},\\
&2)\quad {(\hat p^\mu\hat p_\beta  - \tfrac{1}{3} |\bar \p|^2\Delta^\mu_\beta) \hat q_\alpha   \epsilon^{\alpha\beta\nu\rho}, \text{ and 1 perm}.}
\label{eq:4.12}
\end{align}
Note that a third possible permutations of \Eq{eq:4.12} is not linearly independent of the other two.
\item[d)] one symmetric and traceless three-tensor
\begin{equation}
\hat p^\mu \hat p^\nu \hat p^\rho - \frac{1}{5} |\bar \p|^2 \left[ \hat p^\mu  \Delta^{\nu\rho} + \hat p^\nu  \Delta^{\mu\rho} +\hat p^\rho  \Delta^{\mu\nu}  \right].
\label{eq:4.13}
\end{equation}
\end{enumerate}
Of these only three independent SO(3) representations contribute to $ g^{\mu\nu\rho}_\text{shear}\pi_{\nu\rho}$, namely, those which are symmetric, traceless and orthogonal to fluid velocity in the indecies $\nu$ and $\rho$:
\begin{align}
&g^{\mu\nu\rho}_\text{shear} \pi_{\nu\rho} = \Big\{[p^\nu \pi_{\nu\rho}p^\rho (p^\mu-\bar E_\p u^\mu)-\frac{2}{5}|\bar \p|^2p_\nu\pi^{\nu\mu}] f^\text{shear}_1\nonumber\\
&+ \frac{2}{5}|\bar \p|^2p_\nu\pi^{\nu\mu}f^\text{shear}_2+ \pi_{\nu\rho} p^\nu p^\rho \bar E_\p u^\mu f^\text{shear}_3\Big\}\times \frac{1}{2(e+p)T^2}.\label{eq:pidecomp}
\end{align}
The extension of this procedure to quadratic and higher terms in the Taylor expansions \Eq{eq:taylor1} and \Eq{eq:taylor2} is straightforward, if laborious.

Now we will discuss the transformation rules for the weight functions $f_i$ due to an elementary 2-body decay, \Eq{eq:map2body}. The new vector distribution function is given by the convolution with the decay map
\begin{equation}
g^\mu_b( p,u)=\frac{\nu_a}{\nu_b}\ \int\! \frac{d^3 k}{(2\pi)^3 2 E_\k}  D^a_{b|c}(p^\nu k_\nu)g^\mu_a(k,u).\label{eq:gmunumap}
\end{equation}
Because the Lorentz vectors corresponding to different SO(3) representations in  \Eq{eq:vectordistributionfunction} are orthogonal both before and after the decay, we can use them to project out the desired component and reduce \Eq{eq:gmunumap} to a scalar integral, which can be simplified (the same as \Eq{eq:mapmap}) to a single dimensional integral
\begin{align}
f_i^b(\bar E_\p) &= B \frac{\nu_a}{\nu_b}\frac{m_a^2}{m_b^2 }\frac{1}{2}\int_{-1}^1 dw A_i(w) f_i^a(E(w)).\label{eq:transform}
\end{align}
The calculation is straightforward and the appropriate $A_i(w)$ functions for \Eq{eq:vectordistributionfunction} are
\begin{equation}
A_1^\text{eq} = \frac{Q(w)}{|\bar \p|},\qquad
A_2^\text{eq} = \frac{ E(w)}{\bar E_\p}\\
\end{equation}
where we remind that $E(w)$ and $Q(w)$ were defined as
\begin{subequations}
\begin{align}
E(w) &\equiv \frac{m_a E^a_{b|c}\bar E_\p}{m_b^2}-w \frac{m_a p^a_{b|c}|\bar \p|}{m_b^2},\\
Q(w)&=
\frac{m_a E^a_{b|c}|\bar \p|}{m_b^2}-w \frac{m_a p^a_{b|c}\bar E_\p}{m_b^2}.
\end{align}
\end{subequations}
The same procedure of finding the transformation rules also apply for the irreducible decomposition of  tensor distribution function $g^{\mu\nu}_\text{velocity}$ and $g^{\mu\nu\rho}_\text{shear}$. For the weight functions $f_i$ for $\delta u^\mu$ perturbations in \Eq{eq:dudecomp} we obtain
\begin{align}
A_1^\text{velocity} &= \frac{3}{2}\frac{Q(w)^2}{|\bar \p|^2}-\frac{1}{2}\frac{E(w)^2-m_a^2}{|\bar \p|^2}\quad
\\A_2^\text{velocity} &= \frac{E(w)^2-m_a^2}{|\bar \p|^2},\qquad
A_3^\text{velocity} = A_1^\text{eq}(w)A_2^\text{eq}(w)
\end{align}
while  for the shear-stress case, \Eq{eq:pidecomp},
\begin{align}
A_1^\text{shear} &= \frac{5}{2}\frac{Q(w)^3}{|\bar \p|^3}-\frac{3}{2}\frac{Q(w)}{|\bar \p|}\frac{E(w)^2-m_a^2}{|\bar \p|^2}\\
A_2^\text{shear} &= A_2^\text{velocity}(w)A_1^\text{eq}(w),\quad
A_3 = A_1^\text{velocity}(w)A_2^\text{eq}(w).
\end{align}
Note that some  representations in \Eq{eq:dudecomp} and  \Eq{eq:pidecomp} are just a products of lower dimensional representations and the corresponding functions $A_i(w)$ are equal to the product of their $A_i$'s.  
In summary the \Eq{eq:transform} and functions $A_i(w)$ defines a simple iterative scheme for calculating weight functions $f_i$ for different irreducible components of the vector/tensor distribution function undergoing a 2-body decay \Eq{eq:map2body}, which can be easily extended to a 3-body decay rule by \Eq{eq:3body}. By repeated application of these transformations and summing over all possible decay chains the final decay particle components $f_i$ can be determined. A concrete realization of such scheme is made publicly available \cite{FastReso}.

\section{Decay kernels for azimuthally symmetric and boost-invariant freeze-out surface\label{sec:kernels}}

For azimuthally and boost invariant freeze-out surface
the general decay particle spectrum formula \Eq{eq:fototal} reduces to one dimensional integral 
\Eq{eq:backgroundspectrumintermsofkernels},
where the azimuthal and rapidity integrals are factored out in the 
freeze-out kernels $K_i$. The integral formulas for them are obtained by a straightforward algebra of inserting \Eqs{eq:hypersurfaceelement}, \eq{eq:velocity}, \eq{eq:backgroundshearstress}, and \eq{eq:momentumBjorkenCoord},  in  \Eqs{eq:fototal} and \eq{eq:backgroundSpectrum1}, and collecting terms proportional to temporal and radial components of the freeze-out surface element. Explicitly these are

\begin{equation}
\begin{split}
K^\text{eq}_1(p_T,\bar\chi)  &=  \int_0^{2\pi}\! d\phi \int_{-\infty}^\infty\! d\eta \big\{ f^\text{eq}_1(\bar E_\p) m_T \cosh(\eta)  \\
&+ \left(f^\text{eq}_2(\bar E_\p) - f^\text{eq}_1(\bar E_\p)\right) \bar E_\p \cosh(\bar \chi) \big\}, \\
K^\text{eq}_2(p_T,\bar\chi)  = & \int_0^{2\pi}\! d\phi \int_{-\infty}^\infty\! d\eta \big\{  f^\text{eq}_1(\bar E_\p) p_T \cos(\phi)  \\
&+ \left(f^\text{eq}_2(\bar E_\p) - f^\text{eq}_1(\bar E_\p)\right) \bar E_\p \sinh(\bar \chi) \big\},
\end{split}
\end{equation}
\begin{equation}
\begin{split}
K^\text{shear}_1(p_T,\bar\chi)  &= \int_0^{2\pi}\! d\phi \int_{-\infty}^\infty \!d\eta 
{\Big \{}
\big[f_1^\text{shear}(\bar E_\p) m_T \cosh(\eta)\\
&+ \left(f_3^\text{shear}(\bar E_\p) - f_1^\text{shear}(\bar E_\p)\right) \bar E_\p \cosh(\bar \chi) \big]
\\\times \big[m_T^2 \sinh(\eta)^2 &- \left\{ m_T \sinh(\bar\chi) \cosh(\eta)-p_T \cosh(\bar \chi) \cos(\phi)\right\}^2 \big] \\
 +& \left(f_2^\text{shear}(\bar E_\p) -f_1^\text{shear}(\bar E_\p) \right) \frac{2}{5} (\bar E_\p^2-m^2)  \sinh(\bar \chi)\\
&\times
\left[ m_T \sinh(\bar \chi) \cosh(\eta)  - p_T \cosh(\bar \chi) \cos(\phi) \right]
{\Big \}},
\end{split}
\end{equation}
\begin{equation}
\begin{split}
K^\text{shear}_2(p_T,\bar\chi)  &=  \int_0^{2\pi}\! d\phi \int_{-\infty}^\infty\! d\eta {\Big \{} \big[  f_1^\text{shear}(\bar E_\p) p_T \cos(\phi)\\
& + \left(f_3^\text{shear}(\bar E_\p) - f_1^\text{shear}(\bar E_\p)\right) \bar E_\p \sinh(\bar \chi) \big]
\\\times \big[m_T^2 \sinh(\eta)^2 &- \left\{ m_T \sinh(\bar\chi) \cosh(\eta)-p_T \cosh(\bar \chi) \cos(\phi)\right\}^2 \big] \\
 + &\left(f_2^\text{shear}(\bar E_\p) -f_1^\text{shear}(\bar E_\p) \right) \frac{2}{5} (\bar E_\p^2-m^2) \cosh(\bar \chi)\\
&\times\left[  m_T  \sinh(\bar \chi) \cosh(\eta) - p_T \cosh(\bar \chi) \cos(\phi) \right]
{\Big \}},
\end{split}gg
\end{equation}
\begin{equation}
\begin{split}
K^\text{shear}_3(p_T,\bar\chi)  &=  \int_0^{2\pi}\! d\phi \int_{-\infty}^\infty\! d\eta {\Big \{}  \big[f_1^\text{shear}(\bar E_\p) m_T \cosh(\eta)\\
&+ \left(f_3^\text{shear}(\bar E_\p) - f_1^\text{shear}(\bar E_\p)\right) \bar E_\p \cosh(\bar \chi) \big]
\\\times \big[p_T^2 \sin(\phi)^2 &- \left\{ m_T \sinh(\bar\chi) \cosh(\eta)-p_T \cosh(\bar \chi) \cos(\phi)\right\}^2 \big] \\
 + &\left(f_2^\text{shear}(\bar E_\p) -f_1^\text{shear}(\bar E_\p) \right) \frac{2}{5} (\bar E_\p^2-m^2)\sinh(\bar \chi)\\
& \left[ m_T \sinh(\bar \chi) \cosh(\eta)  - p_T \cosh(\bar \chi) \cos(\phi) \right]
{\Big \}},
\end{split}
\end{equation}
\begin{equation}
\begin{split}
K^\text{shear}_4(p_T,\bar\chi)  &=  \int_0^{2\pi}\! d\phi \int_{-\infty}^\infty\! d\eta {\Big \{}  \big[  f_1^\text{shear}(\bar E_\p) p_T \cos(\phi) \\
&+ \left(f_3^\text{shear}(\bar E_\p) - f_1^\text{shear}(\bar E_\p)\right) \bar E_\p \sinh(\bar \chi)\big]
\\\times \big[p_T^2 \sin(\phi)^2 &- \left\{ m_T \sinh(\bar\chi) \cosh(\eta)-p_T \cosh(\bar \chi) \cos(\phi)\right\}^2 \big] \\
+ &\left(f_2^\text{shear}(\bar E_\p) -f_1^\text{shear}(\bar E_\p) \right) \frac{2}{5} (\bar E_\p^2-m^2) \cosh(\bar \chi) \\
&\left[  m_T  \sinh(\bar \chi) \cosh(\eta) - p_T \cosh(\bar \chi) \cos(\phi) \right]
{\Big \}}.
\end{split}
\end{equation}
\begin{equation}
\begin{split}
K^\text{bulk}_1(p_T,\bar\chi)  = & \int_0^{2\pi}\! d\phi \int_{-\infty}^\infty\! d\eta \big\{ f^\text{bulk}_1(\bar E_\p) m_T \cosh(\eta) \\
& + \left(f^\text{bulk}_2(\bar E_\p) - f^\text{bulk}_1(\bar E_\p)\right) \bar E_\p \cosh(\bar \chi) \big\}, \\
K^\text{bulk}_2(p_T,\bar\chi)  = & \int_0^{2\pi}\! d\phi \int_{-\infty}^\infty\! d\eta \big\{  f^\text{bulk}_1(\bar E_\p) p_T \cos(\phi) \\
& + \left(f^\text{bulk}_2(\bar E_\p) - f^\text{bulk}_1(\bar E_\p)\right) \bar E_\p \sinh(\bar \chi) \big\},
\end{split}
\end{equation}

The above kernels can be evaluated numerically  as two dimensional tables and stored. For a given solution of the hydro equations one can then use \Eq{eq:backgroundspectrumintermsofkernels} to calculate the final particle spectrum.

\clearpage

\bibliography{paper_arxiv_v2}

\end{document}